\documentclass[12pt,a4paper]{article}
\usepackage[latin1]{inputenc}
\usepackage[dvipdf]{epsfig}
\usepackage{bm}
\usepackage{setspace}
\usepackage{cite,graphicx}
\usepackage[scriptsize]{subfigure}

\begin{document}

\title{Homogeneous Gold Catalysis through Relativistic Effects: Addition of Water to Propyne.
\footnote{This work was supported by the Royal Society of New Zealand through a Marsden grant. 
We acknowledge the use of extensive computer time on Massey University's parallel 
supercomputer facilities \emph{Doublehelix} and \emph{Bestgrid}.}}

\author{Matthias Lein, A. Stephen K. Hashmi and Peter Schwerdtfeger\footnote{Dr. Matthias Lein, 
Prof. Dr. Peter Schwerdtfeger, Center for Theoretical Chemistry and Physics, New Zealand Institute for Advanced Study, 
Massey University Auckland, Private Bag 102904, North Shore City, 0745 Auckland, New Zealand; 
Prof. Dr. Stephen K. Hashmi, Organisch-Chemisches Institut, Ruprecht-Karls-Universit\"at Heidelberg, 
Im Neuenheimer Feld 270, D-69120 Heidelberg, Germany.}}

\date{\today}

\vspace{1 cm}

\maketitle
\begin{quote}
\textbf{Keywords:} Gold(III); Homogeneous Catalysis, Density Functional Theory, Relativistic Effects
\end{quote}

\clearpage

For a long time, gold was considered to be catalytically rather inactive due to its 
"chemical inertness", but recent work has demonstrated that gold shows some unexpected 
and quite novel catalytic activities in both heterogeneous \cite{haruta07} and homogeneous systems 
\cite{Hashmi-2008-ChemRev,Hashmi-2004-GoldBull,Hashmi07ChemRev ,Hashmi-2008-AngewChem,Hashmi-2006}, 
with a number of potential commercial applications \cite{goldbook}. This has led to an immense 
increase in research activity in gold catalysis and has been termed the \emph{catalytic gold rush} \cite{Nolan-2007}.

It has been speculated that the unusual catalytic activity in homogeneous gold catalysis is due 
to relativistic effects \cite{Gorin07}. It is now well known and accepted that gold exhibits 
unusually large relativistic effects compared to its neighboring atoms in the periodic table, 
termed the \emph{gold maximum of relativistic effects} \cite{pyykko76,pyykko88}, 
which originates mainly from a very large direct relativistic valence 6$s$-shell contraction 
\cite{baerends90,schwerdtfeger08}. This gives gold rather unique chemical and physical properties 
within the Group 11 elements of the periodic table.
Reviews on this subject were recently given by Pyykk\"{o} \cite{pyykko08} as well as our research 
group \cite{schwerdtfeger08}. In this context we note that relativistic effects substantially 
stabilize the higher oxidation states +III and +V of gold \cite{pyykko88,schwerdtfeger00,kaupp06}, e.g. for the 
corresponding copper and silver compounds only the fluorides are known in the +III oxidation state. 
Nevertheless, beside this important fact it remains to be shown if the unusual catalytic activity 
of Au(III) is indeed due to relativistic effects.

\begin{figure}[h]
  \centering
  \includegraphics[scale=.7]{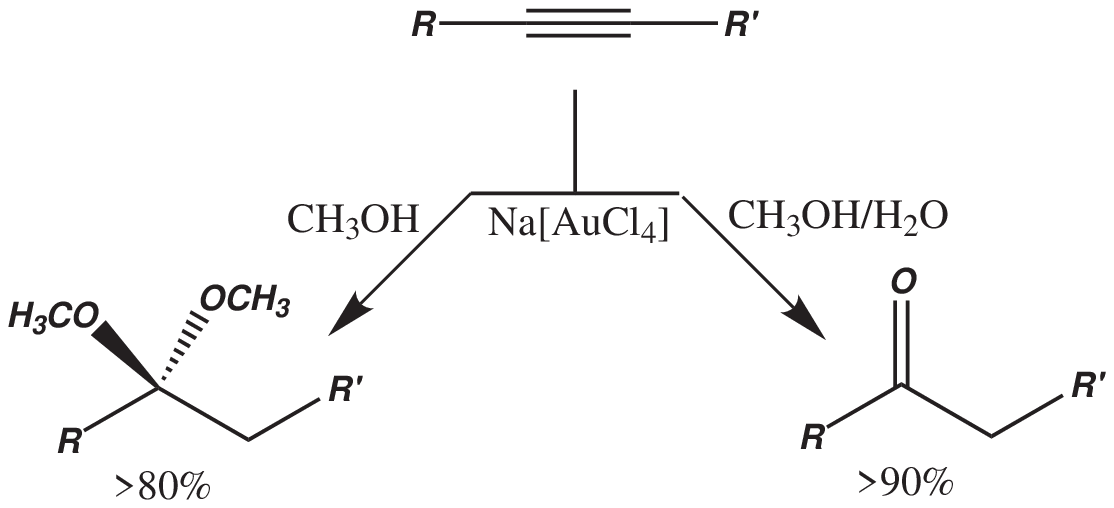}
  \caption{Intermolecular Addition of oxygen-nucleophiles
to alkynes}
  \label{scheme-addition}
\end{figure}

In order to solve this important question, we investigated
the intermolecular addition of
oxygen-nucleophiles to alkynes (see Fig.
\ref{scheme-addition}), which is known for more than fifteen
years now,
\cite{Utimoto-1991-JOrgChem,Utimoto-1991-BullChemSocJpn} and
is part of the standard toolkit
of organic
synthesis\cite{Teles-1998-AngewChem,Hayashi-2002-AngewChem,Hashmi-2003-GoldBull
,Schmidbaur-2004-JMolCatalA}.
This class of reactions has several advantages which makes
it appealing for consideration in a theoretical study
(see computational section). Firstly, the high yield under
mild conditions indicates a clear
thermodynamical preference for the products which enables a
discussion solely on the basis of
the energy differences on the potential energy
hyper-surface. Secondly, the high turnover numbers
and turnover frequencies with a low catalyst loading of only
2 mol\% indicates that the reaction is
likely to be catalyzed by a single catalyst molecule, which
makes the search for the activated complexes feasible.
Third, in the presence of water in the reaction mixture, the
reaction yields the
thermodynamically stable ketone, i.e. the addition product
of water. In order to obtain the product
of the addition of methanol solvent-molecules, i.e. the
acetal, one has to work water-free.
Markovnikov's rule also applies in this case. Finally, one
question that is often raised in the
context of homogeneous gold catalysis concerns the speed of
the reaction. In many cases where
reactions can in principle be catalyzed by a number of
transition metals, gold reacts much
faster than its alternatives\cite{Hashmi-2004-GoldBull}.

\begin{figure}[h]
  \centering
  \includegraphics[scale=0.75]{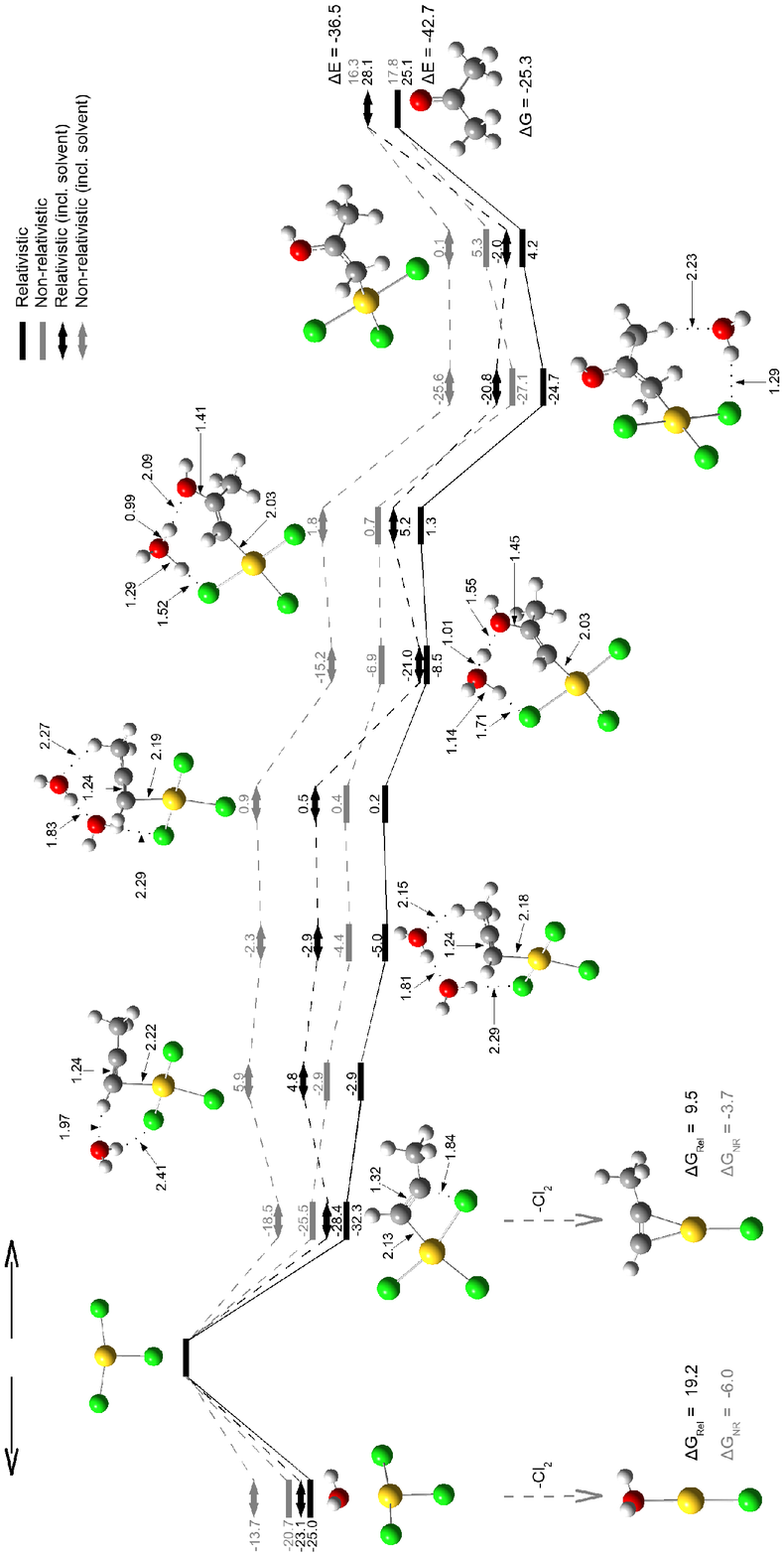}
  \caption{Reaction energy profile of the catalytic gold
process for the
  $\rm H_3C$-C$\equiv$CH + $\rm H_2O \longrightarrow$ $\rm
H_3C$-CO-$\rm CH_3$ reaction.
  All energies in kcal/mol, all distances in \AA. Structures shown are from relativistic 
  gas-phase calculations. For the overall reaction we find: $\Delta$E = -42.7 kcal/mol 
  (gas-phase); $\Delta$E = -36.5 kcal/mol (solvated); $\Delta$G = -25.3 kcal/mol (solvated).}
  \label{fig:energy-profile}
\end{figure}

The addition of water to propyne is catalyzed by $\rm
AuCl_3$ and takes place in several steps
as the overall calculated energy profile in
Fig. \ref{fig:energy-profile} shows. Schr\"oder and Schwarz showed that in the strictly bimolecular regime of the gas-phase "kinetic and entropic restrictions are too large" for the reaction to proceed\cite{Schroeder-2005-InorgChimActa}. Here we show that solvent molecules dramatically change the catalytic cycle and, for the first time explain the mechanistic background of this important mainstay of organic synthesis.

Early studies on reactions of acetylene derivatives with Gold(III) chloride report chlorination of the triple bond and reduction of at least some of the Au(III) to Au(I)\cite{Huettel-1972-ChemBer}. A behavior also observed in the hydrochlorination of acetylene with several gold catalysts\cite{Nkosi-1991-JCatal}. The reduction of Au(III) to Au(I) is highly dependent on the ligand according to Puddephat\cite{Puddephat-1978-GoldBook} which is probably due to Au(I) having a greater affinity for soft polarisable ligands than Au(III). In 2003 Laguna and co-workers studied the catalytic cycle in question and showed that no reduction of the catalyst is taking place under slightly acidic conditions, unambiguously confirming that $\rm AuCl_3$ is indeed the active species in this reaction\cite{Laguna-2003-JACS}.

In the first step of the catalytic cycle the gold moiety coordinates to the triple bond of the alkyne, 
thus occupying the last available coordination site of the Au(III) atom, leading to the well known square 
planar motif common for Au(III) complexes. The bonding situation of the alkyne changes dramatically upon 
coordination. The bond length of the formal triple bond lengthens from 1.21 \AA~in the free propyne to 
1.32 \AA. This change is a clear indication of the activation of the C$\equiv$C bond through the gold 
catalyst. Furthermore, the linear structure of the free propyne is not retained in the $\rm AuCl_3$ 
adduct. This first step of the catalytic cycle is energetically favorable by -32.3 kcal/mol.

In the second step the other reactant enters the scene. A water molecule attaches 
itself to the activated complex, forming two hydrogen bonds. One relatively long 
hydrogen bond (2.41 \AA) between a chlorine atom of the catalyst and the water 
molecule holds the reactant in place from one side while the second hydrogen 
bond between the terminal hydrogen of the alkyne and the water molecule 
clamps the reactant from the other side. The attachment of the water 
molecule to the activated complex is energetically favourable by -2.9 kcal/mol.

A second water molecule comes into play in the next step. This is a necessary requirement in order to bring the activation barriers down. The same mechanism, without the help of a second water molecule, suffers from activation energies more than 20 kcal/mol higher than shown in this investigation. In the case of the second water molecule too, a network of 
hydrogen bonds to hold the reactants in place is formed. Because of the second 
water molecule's steric needs no hydrogen bond can be formed with the terminal 
hydrogen atom of the alkyne. Instead a new hydrogen bond is formed with one of 
the hydrogen atoms of the terminal methyl group at the other end of the substrate 
molecule. The formation of this network of 
relatively weak interactions due to the introduction of a second 
water molecule leads to a stabilization by -5.0 kcal/mol.

The next step leads to the first transition state of this catalytic path. 
Surprisingly, most bond lengths do not change considerably. The hydrogen 
bonds that were present in the preceding step and which hold the two water 
molecules in place are basically the same. This is accompanied by a turning of the 
catalyst along the C-Au-Cl axis by 20\textdegree. This can be viewed as a 
relatively stiff hydrogen bonding network that is being directed along the 
reaction coordinate by the catalyst molecule. This stresses the fact that 
the catalyst does not only activate the triple-bond electronically, but 
acts as a directing agent that guides the necessary reactant molecules into 
the right position for reaction which is consistent with Laguna's findings that at least one chlorine ligand seems to be necessary at the Au(III) center for the catalytic cycle to occur\cite{Laguna-2003-JACS}. This transition state is a mere 0.2 kcal/mol 
energetically above the preceding minimum.

Following the intrinsic reaction channel towards the product leads energetically 
downwards by -8.5 kcal/mol. The resulting structure shows the lead-in to the 
migration of the hydrogen atom from the oxygen atom of the second water molecule 
to the terminal carbon atom and closely resembles the proposed intermediate
from \cite{Laguna-2003-JACS}. However, we are able to show here, that it is not
the nucleophilic attack of an $\rm OH^-$ (which seems an unlikely species in
an acidic solution), but a water molecule that attacks from the back-side.
In this minimum structure the second hydrogen atom of this water molecule is clearly 
starting to dissociate. In this step the role of the first water molecule 
is redefined. In the beginning of the reaction it served as scaffolding to put 
the second water molecule in place through formation of a network of hydrogen 
bonds that connect the catalyst with the substrate through the reactant. 
In this later stage of the reaction the second water molecule also plays 
the role of the mediator, guiding the migration of the hydrogen atom from 
the oxygen atom of the enol to the terminal carbon atom to which the catalyst 
is attached. With the onset of the migration of the hydrogen atom come several 
subtle changes in the structure of the hydrogen bonding network that holds the 
mediating water molecule in place.

The following transition state is again very educt like, but the structural changes 
between the transition state and the preceding minimum are noticeable and subtle. 
First of all, the migrating hydrogen atom is completely abstracted from the enol-oxygen 
atom. At the same time, the bond between the enol-oxygen atom and the carbon atom of the 
now alkyl chain consolidates and the bond length is shortened slightly from 1.45 \AA~ to 
1.41 \AA. This transition state is 1.3 kcal/mol higher in energy than the preceding minimum.

The intrinsic reaction channel leads energetically steeply downwards by -24.7 kcal/mol 
into the product valley. After the migration of the hydrogen atom is completed, the 
mediating water molecule looks for a suitable place, and the terminal methyl group 
seems to be a more favourable target than the terminal hydrogen atoms on the opposite 
side of the alkyl chain which the water molecule preferred in the beginning stages 
of the reaction mechanism. The water molecule forms two hydrogen bonds - one short 
hydrogen bond to the same chlorine atom which guided the water molecule through the 
catalytic cycle, and a longer hydrogen bond with one of the hydrogen atoms 
of the terminal methyl group. The product of this catalytic cycle is now 
obvious in this structure and the enol-form of acetone can easily be seen along with 
the catalyst and one water molecule in this minimum structure.

The last two steps of the catalytic cycle are endothermic by 4.2 kcal/mol and 
25.1 kcal/mol respectively. These amounts of energy are needed to remove the 
remaining water molecule first and then the catalyst second. Note that the relatively 
high energetic cost for removing the catalyst molecule is mitigated by the fact that 
the catalyst can immediately coordinate with another substrate molecule which brings 
it back to the start of the catalytic cycle. The total energetic balance of the reaction is 
$\Delta$E = -42.7 kcal/mol in the gas phase and $\Delta$E = -36.5 kcal/mol if the 
effects of a methanol solution are taken into account. Vibrational and thermal 
corrections to the energy of the solvated system lead to a free energy balance 
of $\Delta$G = -25.3 kcal/mol for the complete catalytic cycle.

We also note that the mechanism of a \emph{cis}-addition proposed by Teles and co-workers \cite{Teles-1998-AngewChem} for the analogue Au(I) system is not reproduced in this study of the Au(III) catalytic cycle. It is indeed sterically impossible for the water molecule to attack the substrate in this fashion as the mediating water molecule acts as a spacer group which pushes the second water molecule into a \emph{trans} approach. It remains to be seen if our results represent a peculiarity of the water addition in the Au(III) case, or if our results can be generalized to Au(I) and Au(III) systems with a variety of oxygen-nucleophiles.

Another reaction that has to be considered in this context is the deactivation 
of the catalyst by a competing reagent in the reaction mixture. Water comes to 
mind since we already know that there are at least trace amounts of water present. 
This reaction is also shown in Fig \ref{fig:energy-profile}. Fortunately, 
we compute the deactivation of the catalyst through a water molecule to be only 
energetically favourable by -23.1 kcal/mol as opposed to -28.4 kcal/mol in the 
competing reaction of the catalyst with the substrate molecule (Table \ref{tab:energies}, 
reactions 2 and 3). Hence, a poisoning of the catalyst by water, which is not observed 
in the experiment, is also not predicted by theoretical consideration.

\begin{table}[h]
\footnotesize{
  \begin{tabular}{rlrclrrl}
& & Reaction & & & $\Delta$E & $\Delta$G\\
 \hline
1 & & CH$_3$C$\equiv$CH + H$_2$O & $\longrightarrow$ & CH$_3$COCH$_3$ & -36.5 & -25.3\\
2 & a & AuCl$_3$ + CH$_3$C$\equiv$CH & $\longrightarrow$ & AuCl$_3$-H$\equiv$CCCH$_3$ & -28.4 & -14.3 & (rel)\\
 & b & AuCl$_3$ + CH$_3$C$\equiv$CH & $\longrightarrow$ & AuCl$_3$-H$\equiv$CCCH$_3$ & -18.5 & -5.7 & (non-rel)\\
3 & a & AuCl$_3$ + H$_2$O & $\longrightarrow$ & AuCl$_3$-H$_2$O & -23.1 & -10.1 & (rel)\\
  & b & AuCl$_3$ + H$_2$O & $\longrightarrow$ & AuCl$_3$-H$_2$O & -13.7 & -2.2 & (non-rel)\\
4 & a & AuCl$_3$-HC$\equiv$CCH$_3$ & $\longrightarrow$ & AuCl-H$\equiv$CCCH$_3$ + Cl$_2$ & 22.2 & 9.5 & (rel)\\
 & b & AuCl$_3$-HC$\equiv$CCH$_3$ & $\longrightarrow$ & AuCl-H$\equiv$CCCH$_3$ + Cl$_2$ & 9.3 & -3.7 & (non-rel)\\
5 & a & AuCl$_3$-H$_2$O & $\longrightarrow$ & AuCl-H$_2$O + Cl$_2$ & 30.4 & 19.2 & (rel)\\
 & b & AuCl$_3$-H$_2$O & $\longrightarrow$ & AuCl-H$_2$O + Cl$_2$ & 5.3 & -6.0 & (non-rel)\\
  \end{tabular}
\caption{\label{tab:energies} Reaction energies and free energies of all computed 
reactions including solvent effects. All values in kcal/mol.}}
\end{table}

An interesting question that we would like to address is the effect that relativity 
has on this catalytic cycle. In order to compare the relativistic and the 
non-relativistic case all computations were repeated with a non-relativistic 
gold atom and the results are shown in Fig \ref{fig:energy-profile} as well. 
The non-relativistic cycle is computed to be much shallower than its 
relativistic counterpart. This indicates a weaker interaction of the 
non-relativistic gold atom with the substrate molecule but it does not 
change the nature of the path or the height of the reaction barriers significantly.
However, a different picture emerges if one considers the dissociation of the 
activated complex into the corresponding Au(I) compound. It has been 
known that relativistic effects stabilise higher oxidation states in 
gold \cite{schwerdtfeger00} and so this reaction lends itself for consideration 
in this study. In order to ascertain whether the Au(III) adducts are stable 
with respect to relativistic effects, the fragmentation reaction into the 
respective Au(I) compounds and dichlorine have been calculated 
(Fig \ref{fig:energy-profile} and Tab \ref{tab:energies}, reactions 4 and 5).
The results of both calculations show, that the relativistic molecules are 
held together much more strongly than the non-relativistic molecules. 
The AuCl$_3$ adducts are less stable by roughly 10 kcal/mol in the 
non-relativistic case. But more importantly, if one adds the vibrational 
and thermal corrections to the energy in order to obtain the free energy 
of the dissociation reaction, one can see that in the relativistic case 
thermodynamics actually prefers to form dichlorine and the Au(I) compound. 
The dissociation reactions are favourable by -3.7 kcal/mol and -6.0 kcal/mol 
respectively. Relativistically this does not happen. Here, the AuCl$_3$ adducts 
are favoured by 9.5 kcal/mol and 19.2 kcal/mol respectively. This shows that 
in a non-relativistic world even the first step of the catalytic cycle is not 
feasible because the propyne-AuCl$_3$ adduct would decompose into the Au(I) 
compound and Cl$_2$. Hence, in this respect the catalytic activity of Au(III) 
compounds is indeed a relativistic effect.

\section{Computational Details}

All elementary steps of the nucleophilic addition of water to propyne catalyzed by 
$\rm AuCl_3$ have been calculated and characterized by using density functional theory 
(DFT) with the Becke-Perdew (BP86) density functional. Correlation consistent 
triple-$\zeta$ basis sets (aug-cc-pVTZ/aug-cc-pVTZ-PP) were used for H\cite{Dunning-1989-JCP}, 
C, O\cite{Dunning-1994-JCP}, Cl\cite{Dunning-1993-JCP} and Au\cite{Peterson-2005-TCA}. 
For Au we used a scalar relativistic energy-consistent small-core pseudopotential 
of the Stuttgart group. To estimate the influence of relativistic effects, 
a non-relativistic pseudopotential \cite{Schwerdtfeger-1989-JCP} with the 
accompanying basis set by Schwerdtfeger and Wesendrup was used\cite{Schwerdtfegerunpub}. 
Solvent effects were taken into account by calculating self-consistent 
polarizable continuum model (PCM) single points at the previously 
obtained structures with methanol as a solvent\cite{Tomasi-1997-JChemPhys,Barone-2002-JChemPhys}. 
Gibbs free energies were obtained by adding vibrational and thermal corrections 
to the result of the PCM single point calculations. The nature of all stationary 
points was examined through the calculation of the second derivative matrix. 
The Gaussian suite of programs was used for all calculations\cite{g03}.

\clearpage

\section*{Table of Contents}

\begin{quote}
 \textbf{\boldmath
 In the catalytic addition of water to propyne the Au(III) catalyst is not stable under 
 non-relativistic conditions and dissociates into a Au(I) compound and Cl$_2$. 
 This implies that one link in the chain of events in the catalytic cycle is broken and 
 relativity may well be seen as the reason why Au(III) compounds are effective catalysts.}
 \end{quote}

\begin{figure}[h]
  \centering
  \includegraphics[scale=0.35]{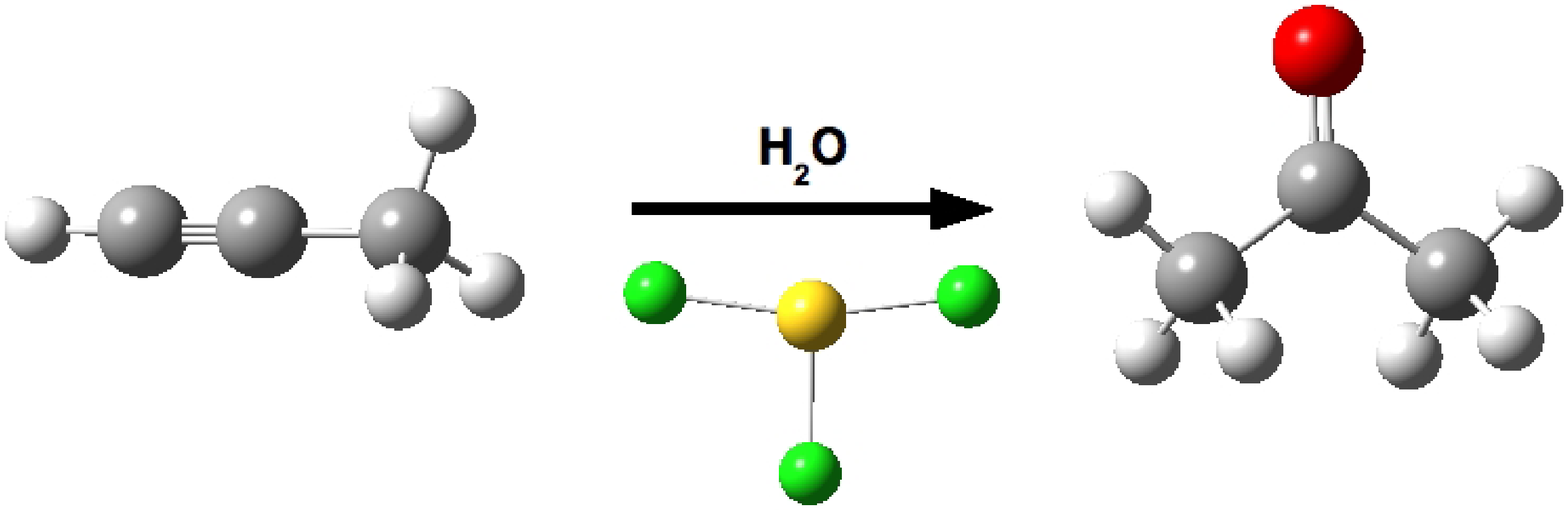}
  \label{fig:TOC}
\end{figure}


\end{document}